\documentclass[lettersize,journal]{IEEEtran}
\usepackage{amsmath,amsfonts}
\usepackage{algorithmic}
\usepackage{array}
\usepackage[caption=false,font=normalsize,labelfont=sf,textfont=sf]{subfig}
\usepackage{textcomp}
\usepackage{stfloats}
\usepackage{url}
\usepackage{verbatim}
\usepackage{graphicx}
\usepackage{cases}
\usepackage{epstopdf}
\usepackage{float}
\usepackage{xcolor}
\def\BibTeX{{\rm B\kern-.05em{\sc i\kern-.025em b}\kern-.08em
    T\kern-.1667em\lower.7ex\hbox{E}\kern-.125emX}}
\usepackage{balance}
\begin{document}
\title{The probabilistic combinatorial attacks on atmospheric continuous-variable quantum secret sharing}
\author{Fangli Yang, \and Liang Chang, \and Minghua Pan
\thanks{Manuscript created April, 2024; This work was supported by the National Natural Science Foundation of China (Grant Nos. U22A2099, 62361021). \textit{(Corresponding author:
Liang Chang).}}
\thanks{F. Yang and L. Chang are with Guangxi Key Laboratory of Trusted Software, Guilin University of Electronic Technology, Guilin 541004, China (email: changl@guet.edu.cn).}
\thanks{M. Pan is with Guangxi Key Laboratory of Cryptography and Information Security, Guilin University of Electronic Technology, Guilin 541004, China.}
}

\markboth{Journal of \LaTeX\ Class Files,~Vol.~, No.~, ~2024}%
{How to Use the IEEEtran \LaTeX \ Templates}

\maketitle

\begin{abstract}
The combination of quantum secret sharing (QSS) and continuous-variable quantum key distribution (CV-QKD) has demonstrated clear advantages and has undergone significant development in recent years. However, research on the practical security of CV-QSS remains limited, particularly in the context of free-space channels, which exhibit considerable flexibility.
In this paper, we study the practical security of free-space CV-QSS, innovatively propose an attack strategy that probabilistically combines two-point distribution attack (TDA) and uniform distribution attack (UDA). We also establish channel parameter models, especially a channel noise model based on local local oscillators (LLO), to further evaluate the key rate. In principle, the analysis can be extended to any number of probabilistic combinations of channel manipulation attacks. The numerical results demonstrate that the probabilistic combination attacks reduce the real key rate of CV-QSS under moderate intensity turbulence, but still enable secure QSS at a distance of 8 km on a scale of hundreds. However, it should be noted that the probabilistic combination attacks will make the deviation between the estimated key rate and the real key rate, i.e., the key rate is overestimated, which may pose a security risk. 

\end{abstract}

\begin{IEEEkeywords}
Quantum secret sharing, Continuous-variable, Free-space channel, Channel manipulation attacks.
\end{IEEEkeywords}

\section{Introduction}
\IEEEPARstart{Q}{uantum} secret sharing (QSS) is a combination of quantum mechanics \cite{bennett2000quantum} and classical secret sharing \cite{1979How,PhysRevLett.95.230505}. A QSS system allows a legitimate user (the dealer) to share a string of secure keys with $n$ participants over an insecure quantum channel. Particularly, in a $(k, n)$-threshold QSS scheme, the dealer splits the secure keys into $n$ parts and distributes them to each participants, requiring no less than $k\leq n$ participants to join forces to determine the string of secure keys. QSS protocols were first proposed for discrete-variable (DV) quantum systems \cite{hillery1999quantum,gottesman2000theory}. Since quantum signals can be effectively prepared, modulated, and measured in quantum optics using continuous-variable (CV) systems, CV-QSS protocols \cite{lau2013quantum,grice2019quantum} were proposed, where the key information is encoded onto the amplitude and phase quadratures of the quantized electromagnetic field of light. Based on the above characteristics, a CV-QSS system has the potential to be easier to implement in practice and has the advantage of being compatible with traditional optical communication networks.

In recent years, CV-QSS has been greatly developed. In Ref. \cite{lau2013quantum}, Lau and Weedbrook proposed a CV-QSS protocol by using continuous-variable cluster states. It is worth noting that this paper is the first to use continuous-variable quantum key distribution (CV-QKD) \cite{Nature421,xu2020secure,yamano2024finite,hajomer2024long} technology to prove the security of CV-QSS. In Ref. \cite{kogias2017unconditional}, Kogias et al. used multi-party entanglement to demonstrate the unconditional security of a CV-QSS system against eavesdroppers in the channel and dishonest participants. However, when the number of participants is large, the preparation of multi-party entangled states becomes a difficult problem. In 2019, Grice and Qi abandoned multiparty entanglement in favor of using weak coherent states to provide easy-to-implement CV-QSS \cite{grice2019quantum}. Therefore, this scheme can also utilize the CV-QKD technique to accomplish the security proof of CV-QSS. Since then, scholars have continuously proposed the CV-QSS protocols based on CV-QKD technology from different angles. Ref. \cite{PhysRevA.101.022301} considered CV-QSS with resources in thermal states and analyzed the finite-size effects of the protocol. Ref. \cite{PhysRevA.103.032410} introduced a CV-QSS scheme using discrete modulated coherent states, which was later extended to a multi-ring discrete modulation CV-QSS \cite{liao2023continuous} with better performance. However, it should be noted that all of the above works are based on fiber channels.

Free-space channels offer significant advantages in terms of infrastructure configuration, facilitating connectivity to moving objects and enabling wider geographical coverage. Consequently, hybrid architectures integrating optical fibers and free-space links are anticipated to assume a pivotal role in facilitating quantum cryptographic communications over extensive networks \cite{ghalaii2023continuous,acosta2024analysis}.
As an important part of quantum cryptographic communication, it is necessary to discuss the free-space architecture of QSS, which is still underdeveloped, especially in the field of continuous variables. In 2021, Ref. \cite{liu2021continuous} presented a CV-QSS protocol based on
thermal terahertz sources in inter-satellite wireless links. In 2023, Ref. \cite{yang2023continuous} analyzed the CV-QSS when the channel transmittance varies according to a uniform probability distribution. Although these two works are based on free-space, they do not discuss in detail some important influencing factors in free-space channels, such as atmospheric turbulence \cite{vasylyev2016atmospheric,PhysRevA.99.053830,trinh2022statistical}, causing beam wandering, beam spreading, etc. 
The primary objective of quantum cryptography is to ensure its practical security. This involves the continuous monitoring of potential attacks. In point-to-point CV-QKD, numerous studies have examined attacks caused by device imperfections, such as LO related attacks \cite{jouguet2013preventing,tan2021wavelength,Shao2022Phase}. Recent research has also investigated channel manipulation attacks \cite{li2018denial,kish2024mitigation}, where Eve manipulates fiber optic channel parameters. Ref. \cite{li2018denial} proposed a denial-of-service attack strategy based on Eve's manipulation of channel transmittance. Building upon this foundation, Ref. \cite{kish2024mitigation} introduced a threat called channel amplification attack in which Eve manipulates the communication channel by amplifying the transmittance. 
This attack has the potential to compromise the security of CV-QKD systems by reducing the key rate, highlighting a significant threat to the system's integrity. However, there is a paucity of discourse within the CV-QSS community concerning such attacks. Given the nature of CV-QSS, involving multiple participants, it is reasonable to infer that channel manipulation could have a more substantial impact compared to CV-QKD.

Based on the above background, we propose the probabilistic combinatorial attacks on free-space quantum secret sharing. The contributions of this paper mainly include the following points:\\
\indent (i) In the CV-QSS, an innovative attack strategy is proposed, which involves the probabilistic combination of two common channel operation attacks, i.e., the TDA and the UDA. The average of the corresponding transmittance model is established, and further formulas for the estimated key rate and the real key rate are given. Theoretically, this analysis method can be extended to any number of probabilistic combinations of channel manipulation attacks.\\
\indent (ii) The free-space channel model is introduced, and in particular, an excess noise model for free-space CV-QSS based on the LLO case is given and minimized. The use of LLO has been demonstrated to prevent the security risk to quantum encryption caused by the transmission of LO through an insecure channel.\\
\indent (iii) The Monte Carlo method is employed to simulate the free-space channel parameters and further analyze the key rate in the finite-size effect and asymptote scenarios. In these scenarios, the modulation variance is optimized and the effects of various parameters on the key rate are analyzed. The numerical results demonstrate that the probabilistic combinatorial attacks reduce the key rate of CV-QSS under moderate intensity turbulence. However, the key rate is still enabled to be secure for quantum secret sharing over a distance of 8 km for hundreds of participants. It is noteworthy that the probabilistic combinatorial attacks result in a discrepancy between the estimated and real key rates, i.e., the key rate is overestimated, which may pose a security risk.

The rest of the paper is organized as follows. In Section \ref{sec:2}, the free-space CV-QSS is described. In Section \ref{sec:3}, we delineate the key rate calculation method for both asymptotic and finite-size cases. In Section \ref{sec:4}, we study the probabilistic combination of the TDA and the UDA. In Section \ref{sec:5}, the free-space channel is modeled in terms of both channel loss and channel noise. The results, including channel parameters and the analysis of security in terms of secret key rate by numerical simulation, are presented in Section \ref{sec:6}. The conclusion is given in Section \ref{sec:7}.

\section{Free-space CV-QSS system description}\label{sec:2}

\subsection{The structure of the CV-QSS protocol}
\begin{figure}[!t]
\centering
\includegraphics[width=3in]{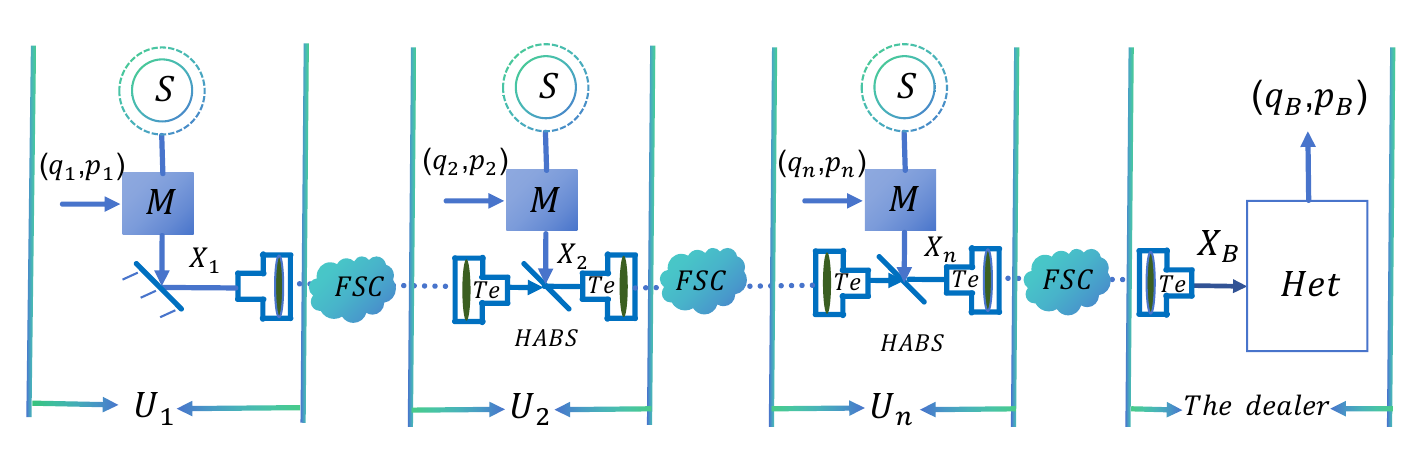}
\caption{The structure of the free-space CV-QSS \cite{yang2023continuous}, comprising a dealer and $n$ participants, denoted as $U_1, U_2,\cdot\cdot\cdot, U_n$. S: the source signal generated by a laser, M: modulator, HABS: highly asymmetric beam splitter, Te: telescope, FSC: free-space channel.}
\label{cvqsslink}
\end{figure}

The structure of the free-space CV-QSS protocol is shown in Fig. \ref{cvqsslink}  \cite{yang2023continuous}, comprising a dealer and $n$ participants, denoted as $U_1, U_2,\cdot\cdot\cdot, U_n$. The procedure of the protocol can be divided into two parts: the quantum stage and the classical post-processing stage.
\subsubsection{\textbf{Quantum stage}}
Each participant $U_j$ $(j=1,\cdot\cdot\cdot,n)$ prepares a local Gaussian modulated quantum state $|\alpha_j\rangle$ described by $X_{j}=X_{j,0}+X_{j,M}+X_{j,T}$ using two random real number $(q_j,p_j)$ from two independent Gaussian distributions of variance $V_M$, where $X_{j,0}$ comes from the quantum fluctuation of the initial coherent state with variance $V_{j,0}=1$, $X_{j,T}$ is the contribution from trusted thermal noise with variance $V_{j,T}$. We assume the variance of each participant is the same as $V_j=V=1+V_M+V_{T}$.

First, $U_1$ sends $|\alpha_1\rangle$ to his (or her) neighbor $U_2$ via a free-space channel (FSC). Next, $U_2$ couples the Gaussian modulated  state to the received signal using a highly asymmetric beam splitter (HABS), and then sends the coupled signal to $U_3$. The remaining participant $U_j$ continues the same process as $U_2$: he (or she) couples the local signal to the received signal from the channel and sends it to the next sparticipant $U_{j+1}$.
In the dealer's side, he (or she) utilizes a telescope (Te) to collect the mixed signal 
and measures it by performing heterodyne detector to obtain the raw data $\{q_B, p_B\}$.
Finally, the above process is repeated several times to generate a sufficiently long set of raw data $D$. 

\subsubsection{\textbf{Classical post-processing stage}}
The dealer estimates the transmittances $\{T_1,T_2,\cdot\cdot\cdot,T_n\}$ by randomly selecting a subset $D_n$ with $n$ pairs from $D$, then randomly picks a pair $\{q_{B},p_{B}\}$ from the remaining data $D/D_n$, and instructs all participants except $U_j$, who is chosen as the honest one, to reveal their corresponding random numbers. 
By utilizing the announced data and $\{T_1,T_2,\cdot\cdot\cdot,T_n\}$, the dealer computes the pair $\{q'_{j},p'_{j}\}$. In this case, a two-party CV-QKD link, denoted as $L_j$, is established between $U_j$ (Alice) and the dealer (Bob). Therefore, we can be able to derive the key rate $r_{j}$ of $L_j$ by using the standard CV-QKD protocol \cite{Nature421} against all the other $n-1$ participants and potential eavesdroppers in the channel. The process is iterated $n$ times to establish a total of $n$ secure CV-QKD links and obtain $n$ secret key rates $\{r_1, r_2, \cdot\cdot\cdot, r_n\}$. Note that in each iteration, a different participant is designated as Alice.
Finally, by performing processes such as error correction and privacy amplification, they use the other, undisclosed subset of data to extract the final security key $k_j$, where $j=1,\cdot\cdot\cdot,n$.
Finally, the dealer encrypts the message $Mess$ with $Mess\oplus \left(k_1\oplus k_2 \oplus\cdot\cdot\cdot\oplus k_n\right)$, thus enabling secret sharing.

\subsection{Parameter estimation}\label{}
In total, the above CV-QSS consists of $n$ local QKD links ($ L_1,\cdot\cdot\cdot, L_n$). In order to evaluate the security of CV-QSS, it is necessary to estimate the main parameters of the channel for each QKD link: the transmittance and the excess noise. In the parameter estimation of CV-QKD link $ L_j$ with the dealer's heterodyne detector efficiency $\eta_e$ and electronic noise $v_{el}$, a normal linear model for $U_j$'s input $X_{j,M}$ and the dealer's output $X_{B}$ is given by
\begin{equation}\label{cm}
X_{B}=t_jX_{j,M}+X_{j,N},
\end{equation}
where $t_j=\sqrt{\frac{\eta_eT_j}{2}}$ and 
$X_{j,N}$ is the aggregated noise with zero mean and variance 
\begin{equation}\label{vn}
V_{j,N}=1+v_{el}+\frac{\eta_eT_j}{2}V_{T}+\frac{\eta_eT_j}{2}\epsilon_j,
\end{equation}
where $\epsilon_j$ is the exess noise of link $ L_j$. Assume that the channel estimation of $L_j$ is made by employing $m$ Gaussian signals, and we define the distributed variables $M_{i}$ and $B_{i}$ $(i\in{1,2,\ldots,m})$ to describe the realizations of the input $X_{j,M}$ and the output $X_{B}$. 
According to Eq. (\ref{cm}), the maximum likelihood estimator of the channel transmittance and channel excess noise are given by  
\begin{equation}\label{t_j}
\hat{t}_j=\frac{\frac{1}{m}\sum_{i=1}^{m}M_{i}B_{i}}{\frac{1}{m}\sum_{i=1}^{m}M^2_{i}}=\frac{E(X_{j,M}X_{B})}{E(X^2_{j,M})},
\end{equation}
\begin{equation}\label{hat{V}_{j,N}}
\begin{split}
\hat{V}_{j,N}&=\frac{1}{m}\sum_{i=1}^{m_0}\left(B_{i}-\hat{t}_jM_{i}\right)^2\\
&=E\left[(X_B-\hat{t}_jX_{j,M})^2\right]\\
&=E(X^2_B)-2\hat{t}_jE(X_BX_{j,M})+\left(\hat{t}_j\right)^2E(X^2_{j,M}).
\end{split}
\end{equation}
Since variables $X_{j,M}$ and $X_{j,N}$ are not correlated, and $X_{j,M}$ follows a Gaussian distribution with a mean of zero and a variance of $V_M$, we can obtain the following equations:
\begin{equation}\label{ebm}
\begin{split}
E(X_{j,M}X_B)&=E\left[X_{j,M}\left(\sqrt{\frac{\eta_eT_j}{2}}X_{j,M}+X_{j,N}\right)\right]\\
&=\sqrt{\frac{\eta_e}{2}}V_ME\left(\sqrt{T_j}\right),
\end{split}
\end{equation}
\begin{equation}\label{X^2_B}
\begin{split}
E(X^2_B)&=E\left(\frac{\eta_eT_j}{2}X^2_{j,M}+X^2_{j,N}\right)\\
&=\frac{\eta_e}{2}E(T_j)\left(V_M+V_T+\epsilon_j\right)+1+v_{el}.
\end{split}
\end{equation}
Substitute Eq. (\ref{ebm}) into Eq. (\ref{t_j}) to get $\hat{t}_j=\sqrt{\frac{\eta_e}{2}}E\left(\sqrt{T_j}\right)$, then the estimator of the channel transmittance can be given by
\begin{equation}\label{hat{T}_j}
\begin{split}
\hat{T}_j=\frac{2(\hat{t}_j)^2}{\eta_e}=\left[E(\sqrt{T_j})\right]^2.
\end{split}
\end{equation}
Similarly, by substituting Eqs. (\ref{ebm}-\ref{hat{T}_j}) into Eq. (\ref{hat{V}_{j,N}}), the estimator of the channel aggregated noise can be rewritten as
\begin{equation}\label{}
\small
\begin{split}
\hat{V}_{j,N}=1+v_{el}+\frac{\eta_e}{2}E(T_j)\left(V_T+V_M+\epsilon_j\right)-\frac{\eta_e}{2}\left[E(\sqrt{T_j})\right]^2V_M.
\end{split}
\end{equation}
According to Eq. (\ref{vn}), we find the estimated value of excess noise $\hat{\epsilon}_j=\left[\hat{V}_{j,N}-(1+v_{el})-\frac{\eta_e}{2}\hat{T}_jV_T\right]\frac{2}{\eta_e\hat{T}_j}$, and by plugging $\hat{T}_j$ and $\hat{V}_{j,N}$ into it, the estimator can be obtained as
\begin{equation}\label{vepj}
\small
\begin{split}
\hat{\epsilon}_j=\frac{E(T_j)}{\left[E(\sqrt{T_j})\right]^2}\left(V_T+V_M+\epsilon_j\right)-\left(V_T+V_M\right).
\end{split}
\end{equation}
We define the variance of the excess noise as $V_{\epsilon_j}=T_j\epsilon_j$, so its estimator is 
\begin{equation}\label{vej}
\small
\begin{split}
\hat{V}_{\epsilon_j}=E(T_j)\left(V_T+V_M+\epsilon_j\right)-\left[E(\sqrt{T_j})\right]^2\left(V_T+V_M\right).
\end{split}
\end{equation}
The practical implementation will introduce additional statistical noise to our estimates due to the finite-size effect. In order to maximize Eve's information from collective attacks, resulting in the lower bound of the key rate in finite-size regime, the worst-case estimators for $U_j$'s each sub-channel where the minimum transmittance $(T_{j})_{min}$ and the maximum excess noise $(V_{\epsilon_j})_{max}$ are taken into account. The two boundaries can be described as
\begin{equation}\label{bound1}
(T_{j})_{min}=\hat{T}_j-Z_\frac{\varepsilon_{PE}}{2}\sigma_{\hat{T_j}},
\end{equation}
and
\begin{equation}\label{bound2}
(V_{\epsilon_j})_{max}=\hat{V}_{\epsilon_j}+Z_\frac{\varepsilon_{PE}}{2}\sigma_{\hat{V}_{\epsilon_j}},
\end{equation}
where $Z_\frac{\varepsilon_{PE}}{2}=6.5$ is a parameter correlated to an error probability of the privacy amplification procedure $\varepsilon_{PE}=10^{-10}$. For the method in \cite{leverrier2010finite,kanitschar2023finite}, the  variance of transmittance $\hat{T}_j$ and excess noise $\hat{V}_{\epsilon_j}$ can be derived as
\begin{equation}
\sigma^2_{\hat{T_j}}=\frac{8}{m}\hat{T}^2_j(1+\frac{\hat{V}_{j,N}}{\eta_e\hat{T}_jV_M})+o(\frac{1}{m^2}),
\end{equation}
\begin{equation}
\sigma^2_{\hat{V}_{\epsilon_j}}=\sigma^2_{\hat{T_j}}V^2_T+\frac{8}{m\eta^2_e}\hat{V}^2_{j,N},
\end{equation}
respectively.

\section{The secret key rate of the protocol}\label{sec:3}
Each QKD link of the CV-QSS will experience a communication interruption with a certain probability due to angle of arrival fluctuations. We assume that the key rate of $L_j$ is $r_{j}$, and the communication interruption probability of $L_j$ is $Pr_j$, where $j=1,2,\cdot\cdot\cdot,n$. The calculation method of $Pr_j$ is described in Appendix A or Ref. \cite{wang2018atmospheric}. Obviously, in order to realize secret sharing, all links must be guaranteed to be non-interruptible, so the non-interruption probability of the whole CV-QSS system is 
\begin{equation}
Pr^{n}_{qss}=\prod \limits_{j=1}^n(1-Pr_j).
\end{equation}
Moreover, to ensure the security of the free-space CV-QSS system, the minimum value in $\{r_1,\cdot \cdot \cdot\,r_n\}$ should be selected as the system key rate. Therefore, the secret key rate of the free-space CV-QSS can be obtained as
\begin{equation}\label{rqss}
K=Pr^{n}_{qss}\times{\rm min}\{r_1,\cdot \cdot \cdot\ ,r_n\}.
\end{equation}

In accordance with the security analysis theory of GMCS CV-QKD \cite{2012Gaussianquantuminformation}, the key rate is closely related to the corresponding channel transmittance and the excess noise. When the original excess noise $\epsilon_0$ introduced by each participant is assumed to be the same, the link with the lowest transmittance among $n$ links is the link with the lowest key rate. The analysis of free-space CV-QKD \cite{wang2018atmospheric} indicates that the channel transmittance decreases with an increase in distance. Consequently, the key rate corresponding to $L_1$, which has the longest distance, will be the minimum key rate among the n links of the CV-QSS. Furthermore, Ref. \cite{yang2023continuous} corroborates this conclusion under the fluctuation channel.
The asymptotic key rate of $L_1$ in the CV-QSS system is given by
\begin{equation}
r_1=\eta I_{A_1B}-\chi_{BE},
\end{equation}
where $I_{A_1B}$ is the Shannon mutual information between $U_1$ and the dealer, and $\chi_{BE}$ is the Holevo quantity of the dealer and Eve. It represents the maximum information that Eve can obtain based on the dealer's variable. 
The Shannon mutual information is calculated by variance $V_{B}$ and the conditional variance
$V_{B|A_1}=1+v_{el}+\frac{\eta_e}{2}V_{\epsilon_{1}}+\frac{\eta_e}{2}T_{1}V_{T}$, 
with the specific calculation formula being
\begin{equation}
I_{A_1B}=\log_2\frac{V_{B}+1}{V_{B|A_1}+1}.
\end{equation}
As for Holevo quantity $\chi_{BE}$, it can be written as \cite{PhysRevA.76.042305}
\begin{equation}\label{chi}
\chi_{ED}=\sum_{m=1}^2G(\lambda_m)-\sum_{m=3}^5G(\lambda_m),
\end{equation}
where $G(\lambda_m)=\frac{\lambda_m+1}{2}\log_2\frac{\lambda_m+1}{2}-\frac{\lambda_m-1}{2}\log_2\frac{\lambda_m-1}{2}$.
The method for calculating symplectic eigenvalues can be referred to in Appendix B of \cite{yang2023continuous}, where it is shown that they depend on the variance $V$, the transmittance $T_{1}$, the channel-added noise 
\begin{equation}
\chi^l_{1}=\frac{1}{T_{1}}-1+\epsilon_{1},
\end{equation}
and the overall noise referred to the channel input \cite{grice2019quantum}
\begin{equation}\label{5}
\chi^t_{1}=\chi^l_{1}+\chi_h/T_{1},
\end{equation}
where $\chi_h=\frac{2-\eta_e+2v_{el}}{\eta_e}$ is the noise caused by the dealer's heterodyne detection. 


It is assumed that the total number of signals transmitted on the free-space channel is $N_0$, where $N_g$ signals are used to generate the key. The finite-size secret key rate between $U_1$ and the dealer can be expressed as
\begin{equation}\label{key2}
\small
R_1=\frac{N_g}{N_0}\left[r_1\left( (T_{1})_{min}, (V_{\epsilon_1})_{max}\right)-\Delta(N_g)\right],
\end{equation}
where $\Delta(N_g)$ is characterized by the speed of convergence of the smooth min-entropy and the security of the privacy amplification \cite{papanastasiou2018gaussian,yang2022finite}. It can be given by
\begin{equation}\label{deta2}
\small
\begin{split}
\Delta(N_g)&\equiv(2dim\mathcal{H}_X+3)\sqrt{\frac{log_{2}(2/\bar{\varepsilon})}{N_g}}\\
&+\frac{2}{N_g}log_{2}(\frac{1}{\varepsilon_{PA}}),
\end{split}
\end{equation}
where $\mathcal{H}_X$ is the Hilbert space and $\bar{\varepsilon}$ is the smoothing parameter.

\section{Probabilistic combination of channel manipulation attacks}\label{sec:4}
Parameter estimation is an important step in CV-QSS protocol, which provides the basis for evaluating key rate in security analysis. The eavesdropper, Eve, has the ability to manipulate the characteristics of the quantum channel and alter its transmittance at will. This can significantly impact estimated parameters by introducing substantial deviations. In this context, we consider that Eve can probabilistically combine a TDA and a UDA.

The channel transmittance of link $L_1$ in the CV-QSS can be decomposed into three constituent parts: $ T_{1,1}$, $T_{1,2}$, and $T_{1,3}$. It is assumed that the susceptibility to a TDA affects the first component, where Eve manipulates the channel transmittance to fluctuate between zero and $T_{1,1}$ according to a two-point distribution of $Y_{1,1}\sim B(1,p)$. The second component is susceptible to a UDA, with the channel transmittance following a uniform distribution of $T_{1,2}Y_{1,2}$ where $Y_{1,2}\sim U(\mu,1)$. Moreover, assuming that the probability of success of the two attacks are $p_{t}$ and $p_{u}$, respectively. The third component remains unaffected by either of these types of attacks. It should be noted that the value range of all parameters $p$, $\mu$, $p_{t}$, and $p_{u}$ is $[0,1]$.

There are four potential scenarios for Eve attacks: two attacks are successfully executed, only a single TDA is successfully executed, only a single UDA is successfully executed, and neither attack is successfully executed. The subsequent relevant parameters are denoted by the subscripts $tu$, $ot$, $ou$, and $nut$, respectively. Then we obtain the corresponding success probabilities $p_{tu}=p_{t}p_{u}$, $p_{ou}=p_{u}(1-p_{t})$, $p_{ot}=p_{t}(1-p_{u})$, and $p_{ntu}=1-p_{t}p_{u}-(1-p_{t})p_{u}-p_{t}(1-p_{u})$. The channel transmittance corresponding to the four cases is 
\begin{equation}
\begin{aligned}
&T_{1,tu}=Y_{1,1}Y_{1,2}T_{1,1}T_{1,2}T_{1,3},\\
&T_{1,ou}=Y_{1,2}T_{1,1}T_{1,2}T_{1,3},\\
&T_{1,ot}=Y_{1,1}T_{1,1}T_{1,2}T_{1,3},\\
&T_{1,ntu}=T_{1,1}T_{1,2}T_{1,3}.
\end{aligned}
\end{equation}
Since we have $E\left(\sqrt{Y_{1,1}}\right)=E\left(Y_{1,1}\right)=p$, $E\left(\sqrt{Y_{1,2}}\right)=\frac{2\left(\mu+\sqrt{\mu}+1\right)}{3(\sqrt{\mu}+1)}$, $E\left(Y_{1,2}\right)=\frac{\mu+1}{2}$, and the variables are independent of each other, then the expected values become
\begin{equation}\label{tjfour}
\begin{aligned}
&E\left(T_{1,tu}\right)=\frac{(\mu+1)p}{2}E\left(T_{1,0}\right),\\
&E\left(T_{1,ou}\right)=\frac{\mu+1}{2}E\left(T_{1,0}\right),\\
&E\left(T_{1,ot}\right)=pE\left(T_{1,0}\right),\\
&E\left(T_{1,ntu}\right)=E\left(T_{1,0}\right),
\end{aligned}
\end{equation}
and
\begin{equation}\label{stjfour}
\begin{aligned}
&E\left(\sqrt{T_{1,tu}}\right)=\frac{2p\left(\mu+\sqrt{\mu}+1\right)}{3(\sqrt{\mu}+1)}E\left(\sqrt{T_{1,0}}\right),\\
&E\left(\sqrt{T_{1,ou}}\right)=\frac{2\left(\mu+\sqrt{\mu}+1\right)}{3(\sqrt{\mu}+1)}E\left(\sqrt{T_{1,0}}\right),\\
&E\left(\sqrt{T_{1,ot}}\right)=pE\left(\sqrt{T_{1,0}}\right),\\
&E\left(\sqrt{T_{1,ntu}}\right)=E\left(\sqrt{T_{1,0}}\right),
\end{aligned}
\end{equation}
where $T_{1,0}=T_{1,1}T_{1,2}T_{1,3}$. In the event that the protocol is unable to ascertain the specific type of channel attack and the corresponding probability, the estimated values of the channel parameters are the average probability of the four cases, i.e.,
\begin{equation}\label{cmostj}
\begin{split}
\small
E\left(\sqrt{T_{1,c}}\right)&=p_{tu}E\left(\sqrt{T_{1,tu}}\right)+p_{ou}E\left(\sqrt{T_{1,ou}}\right)\\&+p_{ot}E\left(\sqrt{T_{1,ot}}\right)+p_{ntu}E\left(\sqrt{T_{1,ntu}}\right),
\end{split}
\end{equation}
\begin{equation}\label{cmotj}
\begin{split}
\small
E\left(T_{1,c}\right)&=p_{tu}E\left(T_{1,tu}\right)+p_{ou}E\left(T_{1,ou}\right)\\&+p_{ot}E\left(T_{1,ot}\right)+p_{ntu}E\left(T_{1,ntu}\right).
\end{split}
\end{equation}

By substituting Eqs. (\ref{cmostj}) and (\ref{cmotj}) into Eqs. (\ref{hat{T}_j}), (\ref{vepj}) and Eqs. (\ref{vej}), the estimators  $\hat{T}_{1,c}$,  $\hat{\epsilon}_{1,c}$ and $\hat{V}_{\epsilon_{1,c}}$ can be obtained. Therefore, the estimated secret key rate is given by
\begin{equation}\label{keye}
\small
K_{c}=K_{1}\left(\hat{T}_{1,c},\hat{V}_{\epsilon_{1,c}}\right).
\end{equation}
By substituting Eqs. (\ref{tjfour}) and (\ref{stjfour}) into Eqs. (\ref{hat{T}_j}) and Eqs. (\ref{vej}), we can get the estimators of channel parameters $\hat{T}_1$ and $\hat{V}_{\epsilon_{1}}$ in the four scenarios. The real key rate should be a composite of the key rates in the presence of single attack, mixed attack, and no attack, i.e.,
\begin{equation}\label{keyr}
\begin{split}
\small
K_{r}&=p_{tu}K_{1}\left(\hat{T}_{1,tu},\hat{V}_{\epsilon_{1,tu}}\right)+p_{ou}K_{1}\left(\hat{T}_{1,ou},\hat{V}_{\epsilon_{1,ou}}\right)\\
&+p_{ot}K_{1}\left(\hat{T}_{1,ot},\hat{V}_{\epsilon_{1,ot}}\right)+p_{ntu}K_{1}\left(\hat{T}_{1,ntu},\hat{V}_{\epsilon_{1,ntu}}\right).
\end{split}
\end{equation}

When there are $M$ channel manipulation attacks, then Eve possesses $M\choose 0$+$M\choose 1$+$\cdot \cdot \cdot$+$M\choose M$ distinct methods for combining these attacks. Given the probability of success for each individual attack, denoted by $p_i$ $(i = 1,\cdot \cdot \cdot,M)$, the probability corresponding to each combination can be determined. Utilizing the aforementioned analysis method for two combination attacks, the average value of the channel transmittance can be obtained. Subsequently, the estimated key rate and the real key rate can be derived. In other words, the above analysis can be generalized to the case where any number of channel manipulation attacks are probabilistically combined.


\section{Free-space channel modeling}\label{sec:5}

\subsection{Channel loss}\label{methodt}
Channel loss can be defined in terms of the optical transmittance. The transmittance is randomly jittered due to beam wandering, broadening, deformation, and scintillation in the atmospheric turbulence channel. Compared to the negative logarithmic Weibull model, the elliptical beam model  better describes the atmospheric turbulence, and its transmittance probability distribution calculated by deriving the Glauber-Sudarshan P-function \cite{vasylyev2016atmospheric} is closer to the real experimental data. Therefore, in this paper, we use an elliptic model for the simulation of free-space channels. See Appendix \ref{elliptical} for a description of this model and also can refer to Ref. \cite{vasylyev2016atmospheric}. 

In the elliptical beam model, the transmittance can be modeled by
\begin{equation}\label{Tj}
\small
T_1=T_{1,r_0} {\rm exp}\left\{-\left[\frac{r_{1,0}/r}{R\left(\frac{2}{{\rm W_{eff}}(\theta_1-\alpha_1)}\right)}\right]^{Q\left(\frac{2}{{\rm W_{eff}}(\theta_1-\alpha_1)}\right)}\right\},
\end{equation}
where {\small $r_{1,0}=\sqrt{x^2_{1,0}+y^2_{1,0}}$, $r$ is the receiving aperture radius, $T_{1,r_0}$} is the transmittance for the centered beam ($r_{1,0}=0$), and {\small ${\rm W_{eff}}(\cdot)$} is the effective squared spot radius. Appendix \ref{ellipticalT_1} shows the derivation of $T_{1,r_0}$ and {\small ${\rm W_{eff}}(\cdot)$}.

Based on the distributions of $\theta_j$ and $\textbf{w}$ (See Appendix \ref{elliptical} for details), the probability density function (PDF) of $T_1$ can be estimated by Monte Carlo simulations. 

\begin{table}
\caption{Default parameters in simulations}
\label{table1}
\centering
\setlength{\tabcolsep}{3pt}
\begin{tabular}{|p{25pt}|p{120pt}|p{80pt}|}
\hline
Symbol& 
Quantity& 
Value 
\\
\hline
 $\lambda_j$& 
Wavelength of $U_{j}$'s Gaussian beam& 
$1.55\times 10^{-6}$m \\
$W_{0j}$& 
Initial radius of $U_{j}$'s Gaussian beam& 
0.06 m \\
$r$& 
Receiving antenna radius& 
0.1 m \\
$d_{cor}$& 
Diameter of fiber core& 
$9\times 10^{-6}$ m \\
$D_f$& 
Focal length of collecting lens& 
0.22 m \\
$\eta$& 
Reconciliation parameter& 
0.98\\
$\eta_e$& 
The efficiency of the dealer's detector& 
0.5\\
$T_H$& 
The transmissivity of the HABS& 
0.99\\
$\epsilon_{0}$& 
Original excess noise introduced by each participant& 
0.01 SNU\\
$v_{el}$& 
The noise variance of the dealer's detector& 
0.1 SNU\\
$V_T$& 
$U_{j}$'s thermal noise& 
0.01 SNU\\
\hline
\end{tabular}
\label{tab1}
\end{table}
\subsection{Channel noise}
Coherent detection of quantum signal pulses requires the use of a high-power LO. In continuous-variable systems, the quantum signal and LO are typically generated by the same laser at the transmitter end and transmitted through a quantum channel, called a transmitted LO (TLO) system. This implementation suffers from security vulnerabilities that can be exploited by eavesdroppers to perform attacks \cite{zhang2024continuous}. In this protocol, we use the LLO \cite{hajomer2024long,qi2015generating} generated by the dealer, thus avoiding the security risk due to the quantum channel transmission. In a free-space LLO CV-QSS system, the total excess noise of can be expressed as
\begin{equation}
\epsilon_1=\epsilon_0+\epsilon_{1,AM}+\epsilon_{1,LE}+\epsilon_{1,LO}+\epsilon_{1,CF},
\end{equation}
where $\epsilon_{1,AM}$ is the modulation noise, which is caused by the imperfection of the modulation device in the preparation of the coherent state. In a CV-QSS system, $n$ participants should prepare coherent states, so the modulation noise consists of $n$ parts. For $L_1$, this noise referred to the channel input can be modeled as
\begin{equation}\label{am}
\begin{split}
\epsilon_{1,AM}=\frac{1}{T_1}\sum_{i=1}^n \left(T_i|\alpha_{smax,i}|^210^{-0.1d_{dB,i}}\right),
\end{split}
\end{equation}
where $|\alpha_{smax,i}|^2\approx 10V_{M}$ is the maximal amplitude of the $U_i's$ signal pulse, and $d_{dB,i}$ is the ratio between the maximal and minimal amplitudes that $U_1$ can output  \cite{marie2017self,shen2023experimental}. $\epsilon_{1,LE}$ is a photon-leakage noise caused by the leakage from the phase reference pulse to the signal pulse \cite{shao2022phase}. 
For $L_1$ of CV-QSS, the phase reference of $U_1's$ signal is coupled to all signal pulses from $U_1$ to $U_n$, that is, the $n$ modulated signals may be contaminated by the phase reference of $U_1's$ signal. Therefore, the photon-leakage noise of $L_1$ in the CV-QSS can be identified as
\begin{equation}\label{le}
\epsilon_{1,LE}=\frac{2E_{R,1}^2}{T_1}\sum_{i=1}^n \left(T_{i}10^{-0.1(R_{e,i}+R_{p,i})}\right),
\end{equation}
where $E_{R,1}$ is the amplitude of the phase reference on dealer's side, $R_{e,i}$ and $R_{p,i}$ are the finite extinction ratios of the amplitude modulator and the polarization beam splitter, respectively.
$\epsilon_{1,LO}$ is the LO noise caused by phase errors, which is given by \cite{marie2017self}
\begin{equation}\label{5}
\epsilon_{1,LO}=2V_{M}(1-e^{-\frac{V_{1,e}}{2}}),
\end{equation}
where $V_{1,e}=V_{1,p}+V_{1,t}+V_{1,m}$ is the variance of the phase noise, which is mainly derived from the phase drift of signal pulse and phase reference in three stages of preparation, transmission  and measurement. We have $V_{1,p}=0$ and $V_{1,t}=0$, when let signal pulse and phase reference be generated from the same optical wave front and transmitted in the same quantum channel \cite{PhysRevA.104.032608}. Therefore, the LO noise mainly comes from phase errors $V_{1,m}$ in the heterodyne detection. In low $V_{1,m}$, the LO noise can be simplified to
\begin{equation}\label{lo}
\epsilon_{1,LO}=V_{M}V_{1,m}=V_{M}\frac{\chi_1+1}{E^2_{R,1}},
\end{equation}
where $\chi_1=\frac{1}{T_1}-1+\epsilon_0+\frac{2-\eta_e+2v_{el}}{\eta_e T_1}$ is the total noise  imposed on the phase-reference. 
From Eqs. (\ref{le}) and (\ref{lo}), $\epsilon_{1,LE}+\epsilon_{1,LO}$ exhibits an increasing trend before undergoing a decrease in relation to $E^2_{R,1}$. This behavior suggests the presence of a minimum value that is attained when $E^2_{R,1}$ satisfies 
\begin{equation}\label{5}
E^2_{R,1}=\sqrt{\frac{T_1 V_{M}(\chi_1(T_1)+1)}{2\sum_{i=1}^n \left( T_{i} 10^{-0.1(R_{e,i}+R_{p,i})}\right)}}.
\end{equation}

The fluctuation noise $\epsilon_{1,CF}={\rm var}\left(\sqrt{T_{1}}\right)V_{M}$ is caused by transmittance fluctuation in a free-space channel, where ${\rm var}\left(\sqrt{T_{1}}\right)=\langle T_{1}\rangle-\langle\sqrt{T_{1}}\rangle^2$ is the variance of the transmittance, which is indicative of the magnitude of the transmittance fluctuations. Note that $\epsilon_{1,AM}$ and $\epsilon_{1,LE}$ are related to the transmittance of other links, and the transmittance $T_i$ and its expectation $\langle T_i\rangle$ of $L_i$ can be obtained using the method in section \ref{methodt}.
Since each part of the noise is independent, the expectation of the total excess noise of $L_1$ in the CV-QSS can be quantified as
\begin{equation}\label{epsilon1}
\langle\epsilon_1\rangle=\epsilon_0+\langle\epsilon_{1,AM}\rangle+\langle\epsilon_{1,LE}\rangle+\langle\epsilon_{1,LO}\rangle+\langle\epsilon_{1,CF}\rangle.
\end{equation}

Considering the volatility of the channel transmittance, we replace $\epsilon_1$ with $\langle\epsilon_1\rangle$ in all the relevant formulas when performing the key rate calculation.

\section{Simulation Results and Discussion}\label{sec:6}
Based on the theoretical analysis in the previous part of this paper, in this section, the parameters such as free-space transmittance and noise are discussed by using numerical simulation, and then the effect of probabilistic combinatorial attacks on the key rate of CV-QSS in free-space is discussed. The values of the relevant parameters are given in Table 1.
\subsection{Channel parameters}
\begin{figure}[!t]
    \centering
    \subfloat[\small PDFs of different turbulence intensities with transmission distance $L=8$ km.]{\includegraphics[width=3in]{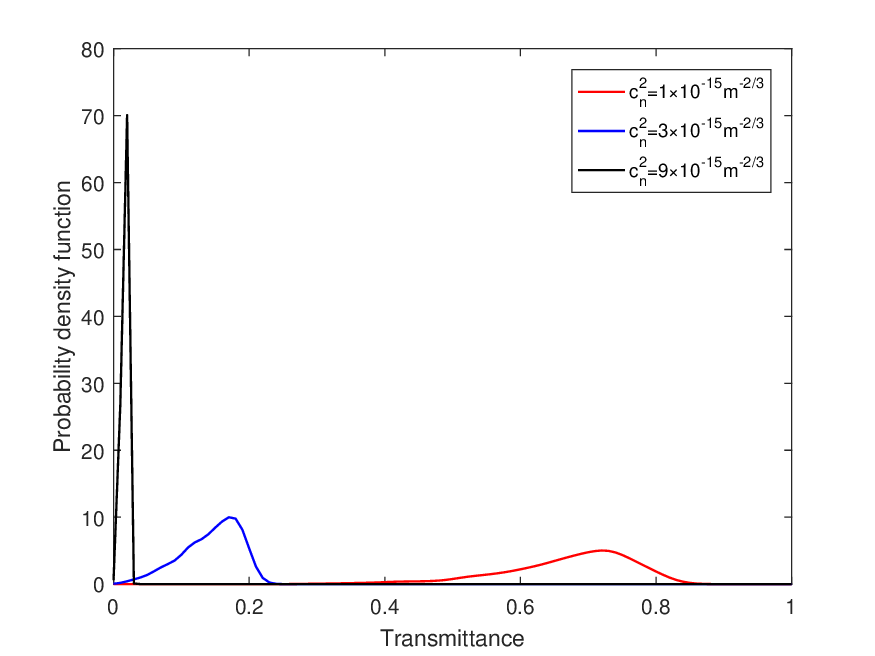}}
    \label{pdfcn1}
    \subfloat[\small PDFs of transmission distances with turbulence intensity $C_n^2=3\times 10^{-15}m^{-2/3}$.]{\includegraphics[width=3in]{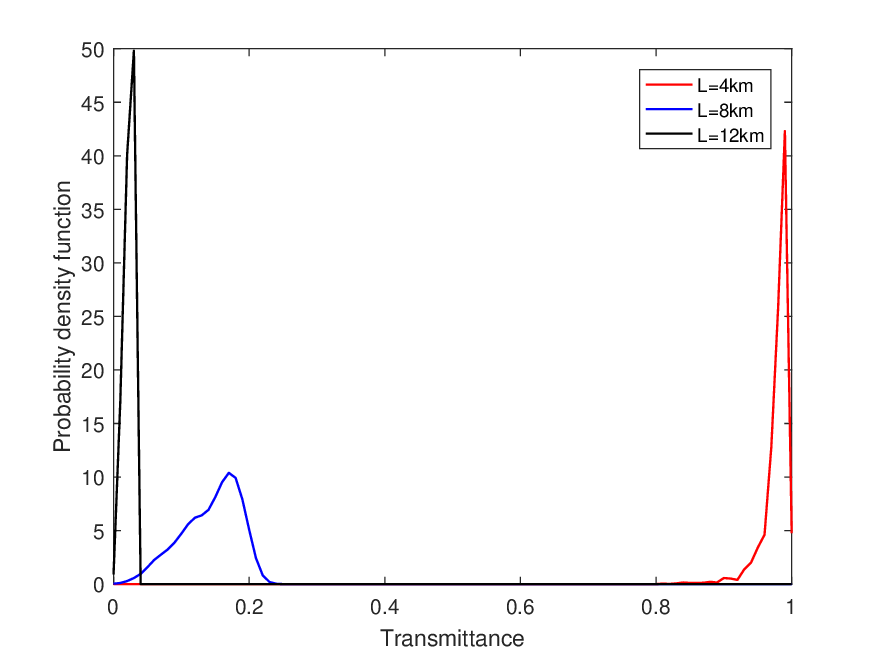}}%
    \label{qkdlink2}
    \caption{ PDFs of the free-space channel transmittance.}
\label{pdf1}
\end{figure}

\begin{figure}[!t]
\centering
\includegraphics[width=3in]{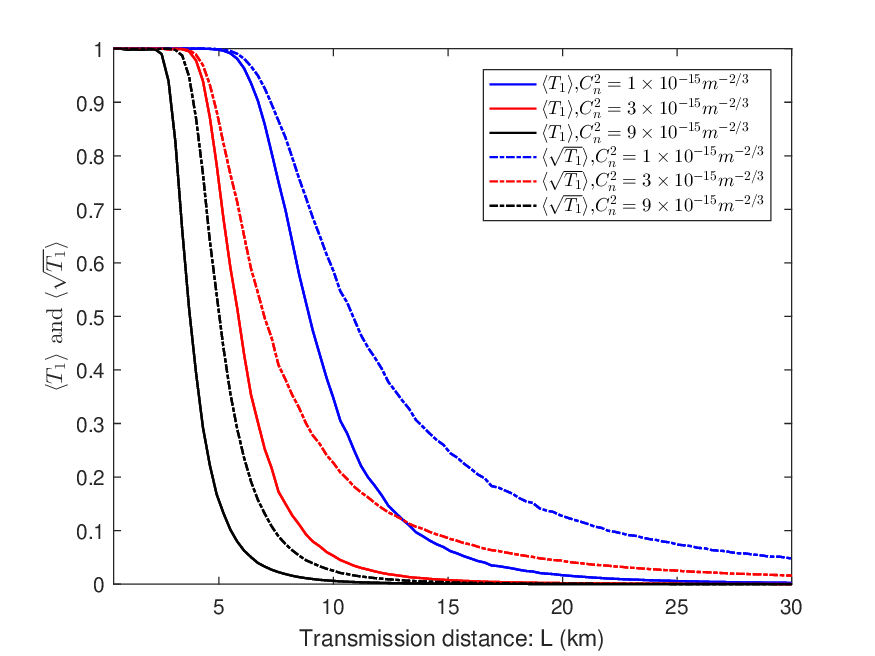}
\caption{The mean values $\langle T_1\rangle$ (solid lines) and $\langle\sqrt{T_1}\rangle$ (dashed lines) as a function of the transmission distance at different turbulence intensities.}
\label{finiteT1}
\end{figure}

The Monte Carlo method is used to generate 1000 random channel transmittances in a free-space channel, which is used to calculate the PDF and the associated channel parameters. Fig. \ref{pdf1} (a) and Fig. \ref{pdf1}(b) show the PDFs of the transmittances at different turbulence intensities and at different transmission distances, respectively. Fig. \ref{finiteT1} shows the mean values $\langle T_1\rangle$ and $\langle\sqrt{T_1}\rangle$ as a function of the transmission distance for different turbulence intensities. From  Fig. \ref{pdf1} and Fig. \ref{finiteT1} it can be seen that as the turbulence intensity and transmission distance increase, the values in the region where the transmittance is centrally distributed and the associated mean values decrease.

\begin{figure}[!t]
\centering
\includegraphics[width=3in]{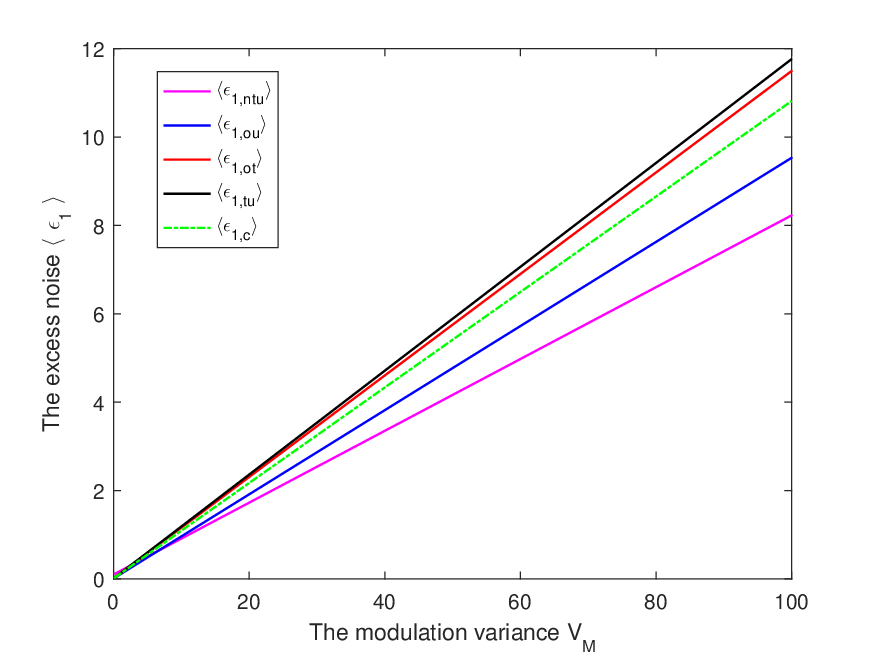}
\caption{The average channel excess noise $\langle\epsilon_{1}\rangle$ as a function of the modulation variance $V_M$, with $L=8km$, $C_n^2=3\times 10^{-15}m^{-2/3}$, $p=0.8$, $\mu=0.3$, $p_{t}=0.7$, and $p_{u}=0.6$.}
\label{averagenoise}
\end{figure}

Fig. \ref{averagenoise} illustrates the average channel excess noise as the modulation variance increases. The coloured solid lines represent the real noise in the four cases where the type of channel attack can be determined, while the dashed lines correspond to the estimated noise when the type of channel attack cannot be determined. As illustrated in the figure, the noise is observed to be at its minimum $\langle\epsilon_{1,ntu}\rangle$ when the channel is not subjected to the TDA and UDA, and the noise is seen to be at its maximum $\langle\epsilon_{1,tu}\rangle$ when it is subjected to a mixture of both of them. This indicates that both attacks introduce noise. Furthermore, the estimation noise is demonstrated to satisfy the inequality $\langle\epsilon_{1,ntu}\rangle<\langle\epsilon_{1,c}\rangle<\langle\epsilon_{1,tu}\rangle$. This observation signifies a discrepancy between the estimated and real noise levels, which in turn leads to a deviation in the subsequent key rate. 

\begin{figure}[!t]
\centering
\includegraphics[width=3in]{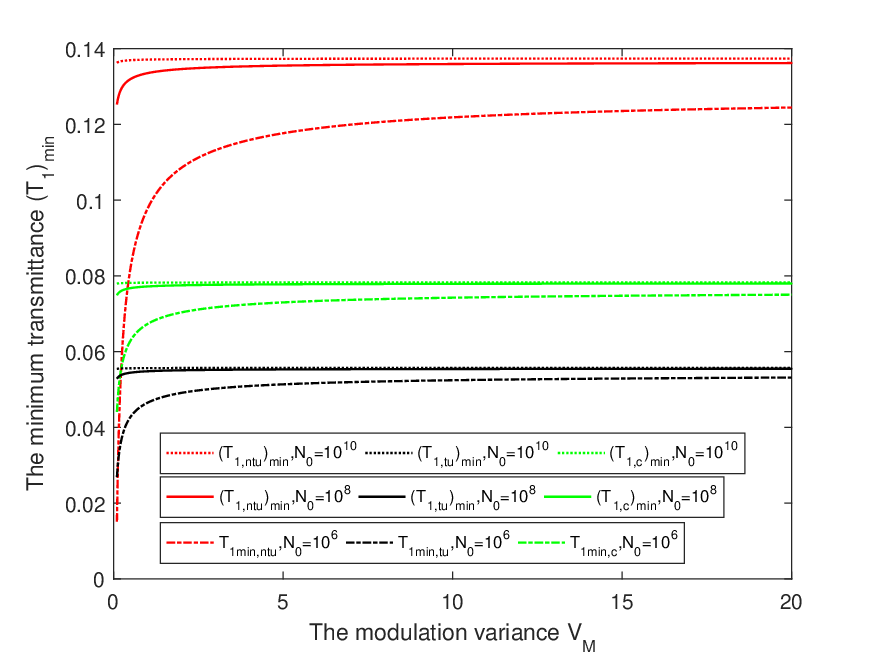}
\caption{The minimum transmittance at different block sizes, with $L=8km$, $C_n^2=3\times 10^{-15}m^{-2/3}$, $p=0.8$, $\mu=0.3$, $p_{t}=0.7$, and $p_{u}=0.6$.}
\label{bound1finite}
\end{figure}

\begin{figure}[!t]
\centering
\includegraphics[width=3in]{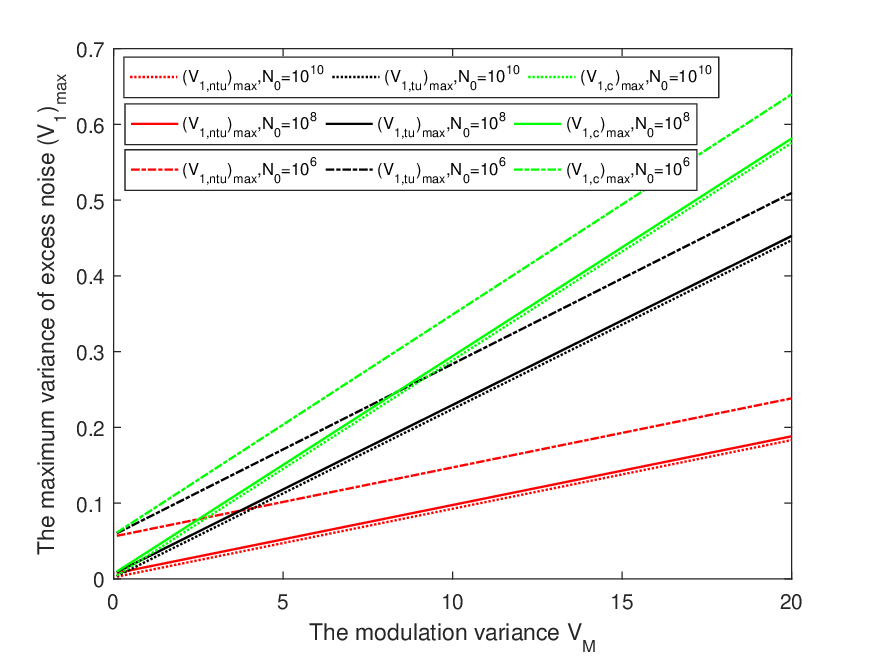}
\caption{The maximum variance of excess noise at different block sizes, with $L=8km$, $C_n^2=3\times 10^{-15}m^{-2/3}$, $p=0.8$, $\mu=0.3$, $p_{t}=0.7$, and $p_{u}=0.6$.}
\label{bound2finite}
\end{figure}

In the context of finite-size effects, the minimum value of transmittance and the maximum value of noise variance can be obtained by utilizing Eqs. (\ref{bound1}) and (\ref{bound2}). Figs. \ref{bound1finite} and \ref{bound2finite} illustrate the impact of block size on these two parameters. The dotted-dashed, solid, and dashed lines in the figures correspond to block sizes of $10^6$, $10^8$, and $10^{10}$, respectively, and the red, black, and green lines represent the cases where the channel is not subject to TDA and UDA, subject to the two types of attacks, and where the type of the attack is not determinable, respectively. From the two figures, it is clear that the larger the block, the larger the minimum value of the corresponding transmittance and the smaller the noise variance. Furthermore, it can be discerned that the modulation parameters exert a negligible influence on the minimum value of the transmittance, with the maximum value of the noise variance being predominantly affected.

\subsection{Secert Key Rate}

\begin{figure}[!t]
\centering
\includegraphics[width=3in]{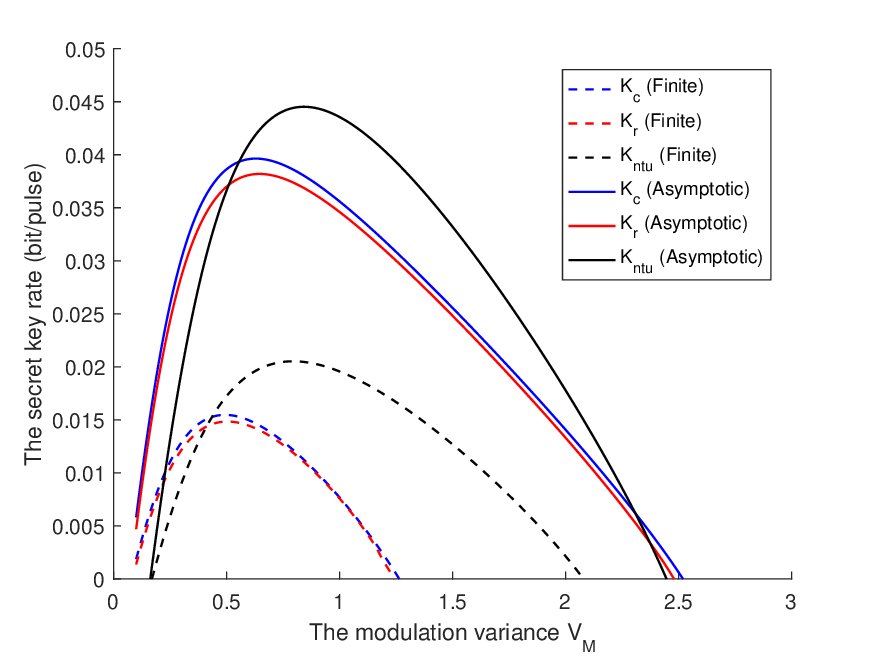}
\caption{The secret key rate as a function of the modulation variance $V_M$, with $L=8km$, $C_n^2=3\times 10^{-15}m^{-2/3}$, $N_0= 10^{10}$, $p=0.8$, $\mu=0.3$, $p_{t}=0.7$, $p_{u}=0.6$, and $n=5$.}
\label{skrmaxV}
\end{figure}

Optimizing the modulation variance is imperative to ensure a high key rate. The plots of key rate with modulation variance for the asymptotic case (dashed lines) and the finite-size case (solid lines) are presented in Fig. \ref{skrmaxV}. As illustrated in the figure, the key rate initially increases with the modulation variance in all cases, attains a maximum value, and subsequently decreases. However, the optimal modulation variance values vary among different cases. To balance the key rate in various cases, we optimize the modulation parameter to $V_M=0.6$ in subsequent numerical simulations.

\begin{figure}[!t]
\centering
\includegraphics[width=3in]{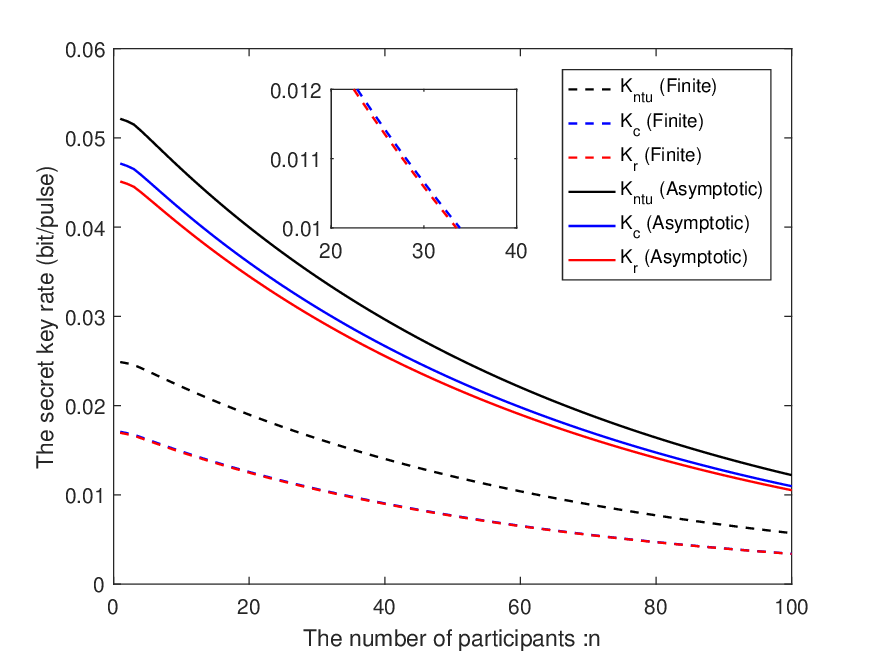}
\caption{The secret key rate as a function of the number of participants $n$, with $L=8km$, $C_n^2=3\times 10^{-15}m^{-2/3}$, $N_0= 10^{10}$, $p=0.8$, $\mu=0.3$, $p_{t}=0.7$, $p_{u}=0.6$, and $V_M=0.6$.}
\label{skrn}
\end{figure}

Fig. \ref{skrn} explores the impact of the number of participants on the key rate. The figure indicates a negative correlation between the number of participants and the key rate, with an increase in participants resulting in a decrease in the key rate under any given scenario. This phenomenon can be attributed to the fact that as the number of participants increases, the excess noise of the system also increases, leading to a reduction in the key rate. It has been observed that $K_r > 0$ when the number of participants reaches 100, although $K_r < K_c < K_{ntu}$. This suggests that the key rate of CV-QSS in free-space channels with moderate turbulence intensity ($C_n^2=3\times 10^{-15}m^{-2/3}$) is affected by channel attacks. However, secure quantum secret sharing over 8 km distances at hundreds of scales can still be realized.

\begin{figure}[!t]
\centering
\includegraphics[width=3in]{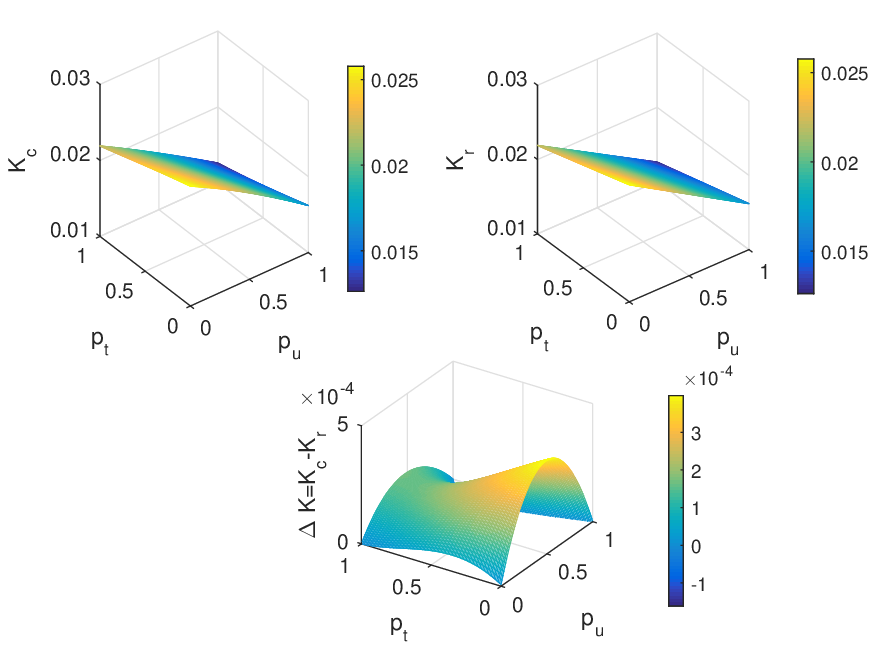}
\caption{The estimated key rate ($K_c$), The real key rate ($K_r$), and the key rate difference ($\Delta K=K_c-K_r$) as a function of the success probabilities of TP attack and UN attack, with $n=5$, $L=8km$, $C_n^2=3\times 10^{-15}m^{-2/3}$, $N_0= 10^{10}$, $p=0.8$, $\mu=0.3$, and $V_M=0.6$.}
\label{skrptpu}
\end{figure}

The subsequent discussion will address the impact of the success probabilities of the TDA and UDA on the key rate. Fig. \ref{skrptpu} demonstrates that as $p_t$ or $p_u$ increases, both the real key rate $K_r$ and the estimated key rate $K_c$ decrease. The difference $\Delta K=K_c-K_r$ between the two key rates varies nonlinearly with the probabilities, yet it is always greater than or equal to zero. This indicates that the attacks not only reduce the security key rate, but also make the deviation between the estimated key rate and the real key rate, that is, the key rate will be overestimated. Therefore, for the security of the CV-QSS system, the average value of the transmittance can be analyzed in conjunction with a machine learning algorithm to obtain the probability of success of the implementation of each attack, and thus the real key rate. The method outlined in Ref. \cite{kish2024mitigation} can be employed to identify the type of the attack by post-processing the data using a decision tree.

\section{Conclusions}\label{sec:7}

In this paper, we presented a novel attack strategy that probabilistically combines two prevalent channel operation attacks (TDA and UDA) in free-space CV-QSS. We established the average of the corresponding transmittance model and derived further formulas for the estimated key rate and the real key rate. Furthermore, the channel noise model based on the LLO case was provided and straightforwardly optimized. Ultimately, the free-space channel parameters and key rate were simulated numerically to optimize the modulation parameters from the perspective of key rate, and the effects of various other parameters, such as the success probabilities of TDA and UDA, on the key rate were analyzed. The numerical results indicated that the probabilistic combinatorial attacks reduce the key rate of CV-QSS under moderate intensity turbulence. However, it enables secure quantum secret sharing at a distance of 8 km for hundreds of scales. It is noteworthy that the probabilistic combinatorial attacks caused a deviation between the estimated key rate and the real key rate, which may introduce security risks. The above results illustrate that if the attacks can be detected and categorized by some methods, and the data can be post-processed to eliminate the security hazards, then secure secret sharing for hundreds of scale participants can be realized in free-space channels. Given that the mean value of the channel transmittance varies with each combination of attacks, future research may focus on detecting and classifying attacks by analyzing the mean value of the transmittance with machine learning algorithms. This approach holds great potential for enhancing the security of free-space CV-QSS.

\begin{appendices}
\section{The communication interruption}\label{sectioninterruption}
In a free-space channel, a large angle-of-arrival fluctuation of the signal can, with a certain probability, lead to an interruption of the quantum communication. Specifically, the beam  jitters randomly in the receiving lens, where case the focus is also randomly distributed. If the focus lies outside the receiving fiber core, the quantum communication is interrupted. Thus, for the QKD between the participant $U_j$ and the dealer ($L_j$), the interruption probability is related to the angle-of-arrival $\theta_{aj}$, fiber core $d_{core}$, and transmission distance $d_j$. We assume that the interruption probability of $L_j$ is $P_j$ and it can be expressed as \cite{wang2018atmospheric}
\begin{equation}
\small
P_j=1-\int_{\frac{{\rm -d_{core}}}{2}}^{\frac{{\rm d_{core}}}{2}}\frac{1}{D_f\sqrt{2\pi \langle\theta_{aj}^2\rangle}}{\rm exp}\left[\frac{-x^2}{2D_f^2\langle\theta_{aj}^2\rangle}\right]{\rm d}x,
\end{equation}
where $D_f$ is the focal length. The variance of $\theta_{aj}$ is
\begin{equation}
\langle\theta_{aj}^2\rangle=\frac{\langle x_{j,0}^2\rangle}{d_j^2},
\end{equation}
where $x_{j,0}$ will be given later in the elliptic model for the channel transmittance analysis. The interruption probability of CV-QSS is
\begin{equation}
P_{QSS}=1-P^{non}_{QSS}=1-\prod \limits_{j=1}^n(1-P_j).
\end{equation}

\section{The elliptical model}\label{elliptical}
The elliptical model assumes that turbulent disturbances in the propagation path cause the Gaussian beam to become elliptical when it reaches the receiver. 


The elliptic beam at the aperture plane of $L_1$ can be characterized by a four-dimensional Gaussian random distribution {\small $\textbf{v}=\{x_{1,0},y_{1,0},W_{1,1},W_{1,2}\}$}, where $(x_{1,0},y_{1,0})$ describes the centroid position of the ellipse, which cause beam wandering, and {\small $W_{1,i}=\sqrt{W_{1,0}^2 {\rm exp}(\phi_{1,i})}(i=1,2)$} are semi-axes of the elliptical spot, which can be used to describe beam broadening and deformation. $W_{1,0}$ is the $U_1$'s Gaussian beam-spot radius and $\phi_{1,i}(i=1,2)$ are variables that conform to normal distributions. The angle $\theta_1\in [0,\pi/2]$ between the longe semi-axis and the $x$ axis is assumed as a uniform distribution. Note that there is no correlation between $\theta_1$ with the other four variables.
The transmittance $T_1$ of $L_1$ in the turbulence channel is related to both a four-dimensional Gaussian random variable $\textbf{w}=\{x_{1,0},y_{1,0},\phi_{1,1},\phi_{1,2}\}$ as well as the random variable $\theta_1$. Variables $x_{1,0}$ and $y_{1,0}$ have no correlations with $\phi_{1,1}$ and $\phi_{1,2}$, while there is a correlation between the latter two variables. $\textbf{w}$ can be described by a covariance matrix
\begin{equation}
\gamma_w=\left(\begin{array}{cccc} 
    \langle x_{1,0}^2\rangle &    0    & 0&0 \\ 
    0 &    \langle y_{1,0}^2\rangle   & 0&0\\ 
    0 & 0 & \langle \phi_{1,1}^2 \rangle &\langle \phi_{1,1}\phi_{1,1} \rangle\\
0 & 0 & \langle \phi_{1,1}\phi_{1,2}\rangle & \langle \phi_{1,2}^2 \rangle
\end{array}\right), 
\end{equation}
with mean value $(0,0,\langle\phi_{1,1}\rangle,\langle\phi_{1,2}\rangle)$, where the diagonal elements of the covariance matrix associated with $x_{1,0}$ and $y_{1,0}$ are given by \cite{PhysRevA.96.043856}
\begin{equation}
\langle x_{1,0}^2\rangle=\langle y_{1,0}^2\rangle=0.33W_{1,0}^2\sigma_{1,1}^2\Omega_1^{-6/7}.
\end{equation}
The symbol $\Omega_1=k_1W^2_{1,0}/2L$ is the Fresnel parameter and 
\begin{equation}
\sigma_{l,1}=1.23C_n^2k_1^{7/6}L^{11/6}
\end{equation}
is the Rytov variance. 
Here $C_n^2$ is the index of refraction structure parameter, and it describes the strength of turbulence. $k_1 = 2\pi/\lambda_1$ is the optical wave number of light with wavelength $\lambda_1$. 
The other covariance matrix elements of $\textbf{w}$ related to variables $\phi_{1,i}$ ($i=1,2$)  are described as
\begin{equation}
\langle \phi_{1,i} \rangle={\rm ln}\frac{(1+2.96\sigma_{l1}^2\Omega_j^{5/6})^2}{\Omega_1^{2}\sqrt{(1+2.96\sigma_{l1}^2\Omega^{5/6})^2+1.2\sigma_{l1}^2\Omega_j^{5/6}}},
\end{equation}
\begin{equation}
\langle \phi_{1,i}^2 \rangle={\rm ln}\left(1+\frac{1.2\sigma_{l1}^2\Omega_1^{5/6}}{(1+2.96\sigma_{l1}^2\Omega_1^{5/6})^2}\right),
\end{equation}
\begin{equation}
\langle \phi_{1,1}\phi_{1,2}\rangle={\rm ln}\left(1-\frac{0.8\sigma_{l1}^2\Omega_1^{5/6}}{(1+2.96\sigma_{l1}^2\Omega_1^{5/6})^2}\right).
\end{equation}

\section{The parameters of $T_{1}$}\label{ellipticalT_1}
We show some details on the elliptic-beam model for $T_{1}$. The maximal transmittance for a centered beam can be given by
\begin{equation}\label{}
\begin{split}
T_{1,r_0}&=1-I_0\left(r^2\left[W_{1,1}^{-2}-W_{1,2}^{-2}\right]\right){\rm exp}^{-r^2\left(W_{1,1}^{-2}+W_{1,2}^{-2}\right)}\\
&-2\left\{1-{\rm exp}\left[-\frac{r^2}{2}\left(W_{1,1}^{-1}-W_{1,2}^{-1}\right)^2\right]\right\}\\
&\times{\rm exp}\left\{-\left[\frac{\frac{(W_{1,1}+W_{1,2})^2}{W_{1,1}^2-W_{1,2}^2}}{R(W_{1,1}^{-1}-W_{1,2}^{-1})}\right]^{Q(W_{1,1}^{-1}-W_{1,2}^{-1})}\right\}
\end{split}
\end{equation}
with the modified Bessel function of i-th order $I_i(\cdot)$, where $R(\cdot)$ and  $Q(\cdot)$ are scale and shape functions, respectively,
\begin{equation}
R(x)=\left[{\rm ln}\left(2\frac{1-{\rm exp}(-r^2x^2/2)}{1-{\rm exp}(-r^2x^2)I_0(r^2x^2)}\right)\right]^{-1/Q(x)},
\end{equation}
\begin{equation}
\begin{split}
Q(x)&=2r^2x^2\frac{{\rm exp}(-r^2x^2)I_1(r^2x^2)}{1-{\rm exp}(-r^2x^2)I_0(r^2x^2)}\\
&\times\left[{\rm ln}\left(2\frac{1-{\rm exp}(-r^2x^2/2)}{1-{\rm exp}(-r^2x^2)I_0(r^2x^2)}\right)\right]^{-1}.
\end{split}
\end{equation}
${\rm W_{eff}}(\cdot)$ is the effective squared spot radius written as
\begin{equation}
{\rm W_{eff}}(x)=2r\left[\textbf{\emph{W}}\left(f_1(x)\frac{4r^2}{W_{1,1}W_{1,2}}f_2(x)\right)\right]^{-\frac{1}{2}},
\end{equation}
where $f_1(x)={\rm exp}[(r^2/W_{1,1}^2)(1+2\cos^2x)]$, $f_2(x)={\rm exp}[(r^2/W_{1,2}^2)(1+2\sin^2x)]$, and $\textbf{\emph{W}}(\cdot)$ is the Lambert {\emph{W}} function \cite{Corless1996}.
\nocite{*}
\end{appendices}


\begin{thebibliography}{1}
\bibliographystyle{IEEEtran}
\bibitem{bennett2000quantum}
C.~H. Bennett and D.~P. DiVincenzo, ``Quantum information and computation,''
  \emph{Nature}, vol. 404, no. 6775, pp. 247--255, 2000.
\bibitem{1979How}
A.~Shamir, ``How to share a secret,'' \emph{Communications of the ACM},
  vol.~22, no.~11, pp. 612--613, 1979.

\bibitem{PhysRevLett.95.230505}
C.~Schmid, P.~Trojek, M.~Bourennane, C.~Kurtsiefer, M.~{\.Z}ukowski, and
  H.~Weinfurter, ``Experimental single qubit quantum secret sharing,''
  \emph{Physical Review Letters}, vol.~95, no.~23, Art. no. 230505, 2005.

\bibitem{hillery1999quantum}
M.~Hillery, V.~Bu{\v{z}}ek, and A.~Berthiaume, ``Quantum secret sharing,''
  \emph{Physical Review A}, vol.~59, no.~3, Art. no. 1829, 1999.

\bibitem{gottesman2000theory}
D.~Gottesman, ``Theory of quantum secret sharing,'' \emph{Physical Review A},
  vol.~61, no.~4, Art. no. 042311, 2000.

\bibitem{lau2013quantum}
H.-K. Lau and C.~Weedbrook, ``Quantum secret sharing with continuous-variable
  cluster states,'' \emph{Physical Review A}, vol.~88, no.~4, Art. no. 042313, 2013.

\bibitem{grice2019quantum}
W.~P. Grice and B.~Qi, ``Quantum secret sharing using weak coherent states,''
  \emph{Physical Review A}, vol. 100, no.~2, Art. no. 022339, 2019.

\bibitem{Nature421}
F.~Grosshans, G.~Van~Assche, J.~Wenger, R.~Brouri, N.~J. Cerf, and P.~Grangier,
  ``Quantum key distribution using gaussian-modulated coherent states,''
  \emph{Nature}, vol. 421, no. 6920, pp. 238--241, 2003.

\bibitem{xu2020secure}
F.~Xu, X.~Ma, Q.~Zhang, H.-K. Lo, and J.-W. Pan, ``Secure quantum key
  distribution with realistic devices,'' \emph{Reviews of Modern Physics},
  vol.~92, no.~2, Art. no. 025002, 2020.

\bibitem{yamano2024finite}
S.~Yamano, T.~Matsuura, Y.~Kuramochi, T.~Sasaki, and M.~Koashi, ``Finite-size
  security proof of binary-modulation continuous-variable quantum key
  distribution using only heterodyne measurement,'' \emph{Physica Scripta},
  vol.~99, no.~2, Art. no. 025115, 2024.

\bibitem{hajomer2024long}
A.~A. Hajomer, I.~Derkach, N.~Jain, H.-M. Chin, U.~L. Andersen, and T.~Gehring,
  ``Long-distance continuous-variable quantum key distribution over 100-km
  fiber with local local oscillator,'' \emph{Science Advances}, vol.~10, no.~1,
  Art. no. eadi9474, 2024.
\bibitem{kogias2017unconditional}
I.~Kogias, Y.~Xiang, Q.~Y. He, and G.~Adesso, ``Unconditional security of
  entanglement-based continuous-variable quantum secret sharing,''
  \emph{Physical Review A}, vol.~95, no.~1, Art. no. 012315, 2017.

\bibitem{PhysRevA.101.022301}
X.~Wu, Y.~Wang, and D.~Huang, ``Passive continuous-variable quantum secret
  sharing using a thermal source,'' \emph{Physical Review A}, vol. 101, no.~2,
  p. 022301, 2020.

\bibitem{PhysRevA.103.032410}
Q.~Liao, H.~Liu, L.~Zhu, and Y.~Guo, ``Quantum secret sharing using discretely
  modulated coherent states,'' \emph{Physical Review A}, vol. 103, no.~3, Art. no.
  032410, 2021.

\bibitem{liao2023continuous}
Q.~Liao, X.~Liu, B.~Ou, and X.~Fu, ``Continuous-variable quantum secret sharing
  based on multi-ring discrete modulation,'' \emph{IEEE Transactions on
  Communications}, vol.~71, no.~10, pp. 6051--6060, 2023.

\bibitem{ghalaii2023continuous}
M.~Ghalaii and S.~Pirandola, ``Continuous-variable
  measurement-device-independent quantum key distribution in free-space
  channels,'' \emph{Physical Review A}, vol. 108, no.~4, Art. no. 042621, 2023.

\bibitem{acosta2024analysis}
V.~M. Acosta, D.~Dequal, M.~Schiavon, A.~Montmerle-Bonnefois, C.~B. Lim, J.-M.
  Conan, and E.~Diamanti, ``Analysis of satellite-to-ground quantum key
  distribution with adaptive optics,'' \emph{New Journal of Physics}, vol.~26,
  no.~2, Art. no. 023039, 2024.

\bibitem{liu2021continuous}
C.~Liu, C.~Zhu, Z.~Li, M.~Nie, H.~Yang, and C.~Pei, ``Continuous-variable
  quantum secret sharing based on thermal terahertz sources in inter-satellite
  wireless links,'' \emph{Entropy}, vol.~23, no.~9, Art. no. 1223, 2021.

\bibitem{yang2023continuous}
F.~Yang, D.~Qiu, and P.~Mateus, ``Continuous-variable quantum secret sharing in
  fast-fluctuating channels,'' \emph{IEEE Transactions on Quantum Engineering},
  vol.~4, no.~01, pp. 1--9, 2023.

\bibitem{vasylyev2016atmospheric}
D.~Vasylyev, A.~Semenov, and W.~Vogel, ``Atmospheric quantum channels with weak
  and strong turbulence,'' \emph{Physical Review Letters}, vol. 117, no.~9, Art. no.
  090501, 2016.

\bibitem{PhysRevA.99.053830}
D.~Vasylyev, W.~Vogel, and F.~Moll, ``Satellite-mediated quantum atmospheric
  links,'' \emph{Physical Review A}, vol.~99, Art. no. 053830, 2019.

\bibitem{trinh2022statistical}
P.~V. Trinh, A.~Carrasco-Casado, H.~Takenaka, M.~Fujiwara, M.~Kitamura,
  M.~Sasaki, and M.~Toyoshima, ``Statistical verifications and deep-learning
  predictions for satellite-to-ground quantum atmospheric channels,''
  \emph{Communications Physics}, vol.~5, no.~1, Art. no. 225, 2022.

\bibitem{huang2013quantum}
J.-Z. Huang, C.~Weedbrook, Z.-Q. Yin, S.~Wang, H.-W. Li, W.~Chen, G.-C. Guo,
  and Z.-F. Han, ``Quantum hacking of a continuous-variable
  quantum-key-distribution system using a wavelength attack,'' \emph{Physical
  Review A—Atomic, Molecular, and Optical Physics}, vol.~87, no.~6, Art. no.
  062329, 2013.

\bibitem{jouguet2013preventing}
P.~Jouguet, S.~Kunz-Jacques, and E.~Diamanti, ``Preventing calibration attacks
  on the local oscillator in continuous-variable quantum key distribution,''
  \emph{Physical Review A—Atomic, Molecular, and Optical Physics}, vol.~87,
  no.~6, Art. no. 062313, 2013.

\bibitem{tan2021wavelength}
X.~Tan, Y.~Guo, L.~Zhang, J.~Huang, J.~Shi, and D.~Huang, ``Wavelength attack
  on atmospheric continuous-variable quantum key distribution,'' \emph{Physical
  Review A}, vol. 103, no.~1, Art. no. 012417, 2021.


\bibitem{Shao2022Phase}
Y.~Shao, Y.~Li, H.~Wang, Y.~Pan, Y.~Pi, Y.~Zhang, W.~Huang, and
B.~Xu, ``Phase-reference-intensity attack on continuous-variable
quantum key distribution with a local oscillator,'' \emph{Physical Review A},
vol. 105, no.~3, Art. no. 032601, 2022.

\bibitem{li2018denial}
Y.~Li, P.~Huang, S.~Wang, T.~Wang, D.~Li, and G.~Zeng, ``A denial-of-service
  attack on fiber-based continuous-variable quantum key distribution,''
  \emph{Physics Letters A}, vol. 382, no.~45, pp. 3253--3261, 2018.

\bibitem{kish2024mitigation}
S.~P. Kish, C.~Thapa, M.~Sayat, H.~Suzuki, J.~Pieprzyk, and S.~Camtepe,
  ``Mitigation of channel tampering attacks in continuous-variable quantum key
  distribution,'' \emph{Physical Review Research}, vol.~6, no.~2, Art. no. 023301,
  2024.

\bibitem{leverrier2010finite}
A.~Leverrier, F.~Grosshans, and P.~Grangier, ``Finite-size analysis of a
  continuous-variable quantum key distribution,'' \emph{Physical Review A},
  vol.~81, no.~6, Art. no. 062343, 2010.

\bibitem{kanitschar2023finite}
F.~Kanitschar, I.~George, J.~Lin, T.~Upadhyaya, and N.~L{\"u}tkenhaus,
  ``Finite-size security for discrete-modulated continuous-variable quantum key
  distribution protocols,'' \emph{PRX Quantum}, vol.~4, no.~4, Art. no. 040306, 2023.

\bibitem{wang2018atmospheric}
S.~Wang, P.~Huang, T.~Wang, and G.~Zeng, ``Atmospheric effects on
  continuous-variable quantum key distribution,'' \emph{New Journal of
  Physics}, vol.~20, no.~8, Art. no. 083037, 2018.

\bibitem{2012Gaussianquantuminformation}
C.~Weedbrook, S.~Pirandola, R.~Garc{\'\i}a-Patr{\'o}n, N.~J. Cerf, T.~C. Ralph,
  J.~H. Shapiro, and S.~Lloyd, ``Gaussian quantum information,'' \emph{Reviews
  of Modern Physics}, vol.~84, no.~2, Art. no. 621, 2012.

\bibitem{PhysRevA.76.042305}
J.~Lodewyck, M.~Bloch, R.~Garc{\'\i}a-Patr{\'o}n, S.~Fossier, E.~Karpov,
  E.~Diamanti, T.~Debuisschert, N.~J. Cerf, R.~Tualle-Brouri, S.~W. McLaughlin
  \emph{et~al.}, ``Quantum key distribution over 25 km with an all-fiber
  continuous-variable system,'' \emph{Physical Review A}, vol.~76, no.~4, Art. no.
  042305, 2007.

\bibitem{papanastasiou2018gaussian}
P.~Papanastasiou, C.~Ottaviani, and S.~Pirandola, ``Gaussian one-way thermal
  quantum cryptography with finite-size effects,'' \emph{Physical Review A},
  vol.~98, no.~3, Art. no. 032314, 2018.

\bibitem{yang2022finite}
F.~Yang, D.~Qiu, L.~Chen, and X.~Wan, ``Finite-size analysis of thermal states
  quantum cryptography with the optimal noise,'' \emph{Annalen der Physik},
  vol. 534, no.~1, Art. no. 2100268, 2022.

\bibitem{zhang2024continuous}
Y.~Zhang, Y.~Bian, Z.~Li, S.~Yu, and H.~Guo, ``Continuous-variable quantum key
  distribution system: Past, present, and future,'' \emph{Applied Physics
  Reviews}, vol.~11, no.~1, Art. no. 011318, 2024.

\bibitem{qi2015generating}
B.~Qi, P.~Lougovski, R.~Pooser, W.~Grice, and M.~Bobrek, ``Generating the local
  oscillator “locally” in continuous-variable quantum key distribution
  based on coherent detection,'' \emph{Physical Review X}, vol.~5, no.~4, Art. no.
  041009, 2015.

\bibitem{marie2017self}
A.~Marie and R.~All{\'e}aume, ``Self-coherent phase reference sharing for
  continuous-variable quantum key distribution,'' \emph{Physical Review A},
  vol.~95, no.~1, Art. no. 012316, 2017.

\bibitem{shen2023experimental}
T.~Shen, X.~Wang, Z.~Chen, H.~Tian, S.~Yu, and H.~Guo, ``Experimental
  demonstration of llo continuous-variable quantum key distribution with
  polarization loss compensation,'' \emph{IEEE Photonics Journal}, vol.~15,
  no.~2, pp. 1--9, 2023.

\bibitem{shao2022phase}
Y.~Shao, Y.~Li, H.~Wang, Y.~Pan, Y.~Pi, Y.~Zhang, W.~Huang, and B.~Xu,
  ``Phase-reference-intensity attack on continuous-variable quantum key
  distribution with a local local oscillator,'' \emph{Physical Review A}, vol.
  105, no.~3, Art. no. 032601, 2022.

\bibitem{PhysRevA.104.032608}
Y.~Shao, H.~Wang, Y.~Pi, W.~Huang, Y.~Li, J.~Liu, J.~Yang, Y.~Zhang, and B.~Xu,
  ``Phase noise model for continuous-variable quantum key distribution using a
  local local oscillator,'' \emph{Physical Review A}, vol. 104, Art. no. 032608, 2021.

\bibitem{PhysRevA.96.043856}
D.~Vasylyev, A.~A. Semenov, W.~Vogel, K.~G\"unthner, A.~Thurn, O.~Bayraktar,
  and C.~Marquardt, ``Free-space quantum links under diverse weather
  conditions,'' \emph{Physical Review A}, vol.~96, Art. no. 043856, 2017.

\bibitem{Corless1996}
R.~Corless, G.~Gonnet, D.~Hare, D.~Jeffrey, and D.~Knuth, ``On the lambertw
  function,'' \emph{Advances In Computational Mathematics}, vol.~5, pp. 329--359, 1996.
%
%
%
%
%
%
%
%
%
%
%
%
%
%
%
%
%
%


\end{thebibliography}
\end{document}